\documentclass[12pt]{article}   

\usepackage{amsmath,amssymb,amsfonts,bm,amsthm}
\usepackage{epsfig}
\usepackage{graphicx,rotating,lscape}
\usepackage{hyperref}
\usepackage{natbib} 
\usepackage{booktabs} 
\usepackage[all]{xy}
\usepackage{color}
\usepackage{latexsym}
\usepackage{float}
\usepackage{xcolor}
\usepackage{soul}
\usepackage{enumitem}
\usepackage{makecell}
\usepackage{tabularx}
\newcolumntype{L}{>{\raggedright\arraybackslash}X}
\newcolumntype{Y}{>{\centering\arraybackslash}X}

\newtheorem{theorem}{Theorem}      
\newtheorem{proposition}{Proposition}
\newtheorem{definition}{Definition}

\newtheorem{remark}{Remark}
\newtheorem{property}{Property}
\theoremstyle{definition}

\title{Surrogate method for partial association between mixed data with application to well-being survey analysis}

\vspace{-0.4in}
\author{Shaobo Li\footnote{
		Shaobo Li is Assistant Professor, University of Kansas School of Business, Lawrence, Kansas, USA (\href{mailto:shaobo.li@ku.edu}{shaobo.li@ku.edu}); 
		Zhaohu Fan is Ph.D. student, Department of Operations, Business Analytics and Information Systems, University of Cincinnati Lindner College of Business, Cincinnati, Ohio, USA (\href{mailto:fanzh@ucmail.uc.edu}{fanzh@ucmail.uc.edu});  
		Ivy Liu is Associate Professor, School of Mathematics and Statistics, Victoria University of Wellington, Wellington, New Zealand (\href{mailto:ivy.liu@vuw.ac.nz}{ivy.liu@vuw.ac.nz}); and 
		Philip S. Morrison is Professor Emeritus, School of Geography, Environment and Earth Sciences, Victoria University of Wellington, Wellington, New Zealand (\href{mailto:philip.morrison@vuw.ac.nz}{philip.morrison@vuw.ac.nz}).}, Zhaohu Fan, Ivy Liu, Philip S. Morrison, Dungang Liu\footnote{Dungang Liu is the corresponding author (\href{mailto:liudg@ucmail.uc.edu}{dungang.liu@uc.edu}) and Associate Professor, Department of Operations, Business Analytics and Information Systems, University of Cincinnati Lindner College of Business, Cincinnati, Ohio, USA.}}

\begin{document}
\renewcommand{\arraystretch}{1.1}
\baselineskip = 8.1mm
\parskip = 4mm

\thispagestyle{empty}
\maketitle


\newpage
\setcounter{page}{1} 
\begin{center}
	\LARGE
Surrogate method for partial association between mixed data with application to well-being survey analysis

\end{center}
  \vspace{1cm}
  \begin{abstract}
  \baselineskip = 6mm
This paper is motivated by the analysis of a survey study of college student well-being before and after the outbreak of the COVID-19 pandemic. A statistical challenge in well-being survey studies lies in that outcome variables are often recorded in different scales, be it continuous, binary, or ordinal. The presence of mixed data complicates the assessment of the associations between them while adjusting for covariates. In our study, of particular interest are the associations between college students' well-being and other mental health measures and how other risk factors moderate these associations during the pandemic. To this end, we propose a unifying framework for studying partial association between mixed data. This is achieved by defining a unified residual using the surrogate method. The idea is to map the residual randomness to the same continuous scale, regardless of the original scales of outcome variables. It applies to virtually all commonly used models for covariate adjustments. We demonstrate the validity of using such defined residuals to assess partial association. In particular, we develop a measure that generalizes classical Kendall's tau in the sense that it can size both partial and marginal associations. 
More importantly, our development advances the theory of the surrogate method developed in recent years by showing that it can be used without requiring outcome variables having a latent variable structure. The use of our method in the well-being survey analysis reveals (i) significant moderation effects (i.e.,  the difference between partial and marginal associations) of some key  risk factors; and (ii) an elevated moderation effect of physical health, loneliness, and accommodation after the onset of COVID-19.

	\vspace{0.3in}
	Keywords: mental health,pandemic,covariate adjustment, Kendall's tau, moderation analysis, partial correlation, surrogate residual.
  \end{abstract}

\newpage
\section{Introduction}\label{sec:intro}


The COVID-19 pandemic, which broke out in January 2020, has brought unprecedented impact to human beings including both physical and mental health. It has imposed an additional strain on college students whose well-being before the pandemic was already a matter of international concern \citep{sales2001academic}.  With the arrival of COVID-19, it has become even more important to study student well-being and mental health \citep{burns2020assessing,lederer2021more}. In so doing it has become equally important to understand the associations between measures of well-being and other dimensions of mental health such as anxiety, psychological distress and life satisfaction \citep{cao2020study,every2020psychological,groarke2020loneliness}. However, the presence of mixed types of outcome data (e.g., binary, ordinal, and continuous), as often seen in other well-being studies, is a well-recognized challenge in association analysis. 

Our study is motivated by the analysis of the YOU Student Well-being Survey (referred to as YOU survey hereafter). The survey data were collected in both pre-pandemic and early pandemic periods, from two cohorts of first year college students attending a major university in New Zealand. Details of data description can be found in Section 2. The goal is to better understand the associations between student's well-being and other dimensions of mental health which are recorded in different data types, and how such associations are moderated by other risk factors before and after the outbreak of the COVID-19 pandemic. To achieve this goal, we propose a unifying framework for studying partial association between mixed types of data with covariates adjustment. 

%


Partial association examines the association between multiple variables with the effect of a set of covariates removed \citep{fisher1924distribution}. To analyze mixed types of outcome data, a large body of the literature treats it as a nuisance (as a parameter or a latent variable) in statistical modeling, where the focus of inference is not on partial correlation but how factors influence the outcome \citep{de2013analysis}. Commonly used modeling methods for mixed data include the latent variable methods  \citep{catalano1992bivariate,fitzmaurice1995regression,sammel1997latent,gueorguieva2001correlated,teixeira2009correlated,zhang2018modeling};   the pseudo/composite likelihood methods \citep{prentice1991estimating,zhao1992multivariate,faes2008high,najita2009novel,bai2020multivariate};
the conditional regression methods \citep{cox1992response,de2007general}; and the copula methods \citep{song2009joint,he2012gaussian,stober2015comorbidity,zilko2016copula,yang2022nonparametric}.

In well-being and mental health research, partial association itself has become of inferential interest. For example, \citet*{zhu2012nonparametric} and \citet*{jiang2014identifying} proposed nonparametric/semiparametric approaches for testing association between mixed variables while adjusting for covariates. Nevertheless, their inferences are limited to hypothesis testing, i.e., testing whether or not partial association is zero. The resulting $p$-value provides a dichotomous answer (yes/no), but it does not describe the effect size nor reveal the direction of partial association. In our well-being study on college students, a question is: to what extent do the factors $\bm X$ (e.g., physical health and accommodation) contribute to moderating the association between {\it well-being} ($Y_{W}$) and a measure of mental health, say, {\it anxiety} ($Y_{A}$)? To answer this question, an effect size measure of partial association is a necessity to quantify the association reduction after adjusting for the covariates under investigation.

A lingering technical challenge in well-being studies is that the outcome variables are often reported in different scales. In our study, for instance, {\it well-being} ($Y_{W}$) is a continuous score derived from a World Health Organization survey \footnote{\url{https://www.psychcongress.com/saundras-corner/scales-screeners/well-being-index/who-five-well-being-index-who-5}}, whereas {\it anxiety} ($Y_{A}$) is an ordered categorical variable whose value is recorded using an ordinal scale from 1 (strongly disagree) to 5 (strongly agree) for the survey question: ``I feel nervous, anxious or on an edge''. To measure the size of moderation effect, a traditional approach is to consider two regression models $Y_{W}=\alpha+\beta Y_{A}+\epsilon$ and  $Y_{W}=\alpha'+\beta' Y_{A}+\bm \gamma^T \bm X + \epsilon'$, without and with the covariates $\bm X$. The coefficient estimates $\hat{\beta}$ and $\hat{\beta'}$ measure marginal and partial associations, respectively. Their difference reflects the moderation effect \citep{mackinnon2000equivalence}. A change of 10\% is the rule of thumb used to determine clinical significance \citep{greenland2008invited,lee2014cutoff}. Although this regression coefficient method is used widely, our survey analysis shows that it yields results that are inconsistent, self-contradictory, and therefore confusing to domain researchers. 

\begin{table}
	\centering
	\caption{Numerical changes of regression coefficients before and after covariate adjustments in the analysis of the 2020 student cohort data. Panel A shows the result when {\it well-being} is treated as the response and {\it anxiety} as a predictor in a regression model. Panel B and C show the results when {\it anxiety} is treated as the response with different regression models.} \label{tab:ex1}
	\begin{tabular}{lccccc}
		\hline
		\multicolumn{6}{l}{Panel A: {\it well-being} is treated as the response variable (linear model)}\\
		\hline
		& $Y_{W} \sim Y_{A}$ && $Y_{W} \sim Y_{A}+ \bm X$ && \% change of coefficients\\
		\cline{2-2} \cline{4-4} \cline{6-6}
		\textit{anxiety-level 2} & $-4.588^{**}$   && $-2.180$   && $-52.5$\\ 
		& (2.290)		  && (2.172) 		&&\\
		\textit{anxiety-level 3} & $-11.039^{***}$ && $-7.341^{***}$ && $-33.5$\\
		& (2.251)		  && (2.150)		&&\\
		\textit{anxiety-level 4} & $-21.021^{***}$ && $-15.526^{***}$ && $-26.1$\\
		& (2.204)		  && (2.134)		&&\\
		\textit{anxiety-level 5} & $-31.623^{***}$ && $-23.466^{***}$ && $-25.8$\\
		& (2.600)		  && (2.549)		&&\\
		\hline
		Average &&&&& $\mathbf -34.5$\\
		\hline
		\multicolumn{6}{l}{Panel B: {\it anxiety} is treated as the response variable (adjacent category logit model)}\\
		\hline
		& $Y_{A} \sim Y_{W}$ && $Y_{A} \sim Y_{W}+ \bm X$ && \% change of coefficients\\
		\cline{2-2} \cline{4-4} \cline{6-6}
		\textit{{\it well-being}} & $-0.032^{***}$ && $-0.029^{***}$ && $-9.3$\\
		& (0.002)		 && (0.002)		&&\\
		\hline
		\multicolumn{6}{l}{Panel C: {\it anxiety} is treated as the response variable (stereotype model)}\\
		\hline
		\textit{{\it well-being}} & $0.124^{***}$ && $0.118^{***}$ && $-4.6$\\
		& (0.012)		 && (0.011)		&&\\
		\hline
		\textit{Note:} & \multicolumn{5}{r}{$^{*} p<0.1$; $^{**} p<0.05$; $^{***} p<0.01$}
	\end{tabular}
\end{table}

To quantify the association between {\it well-being} and {\it anxiety}, the regression coefficient method forces us to pick one of the two as the response and the other as a predictor even when the association may be bi-directional. The analysis result is presented in Panel A of Table \ref{tab:ex1} if {\it well-being} is treated as the ``response'' variable in a regression model. The last column shows the percentage change of the estimated coefficients of $Y_{A}$ in the models with and without the covariates $\bm X$, which are shown in the 2nd and 3rd columns. Since $Y_{A}$ is a categorical variable with five levels, the models have four dummy variables with four coefficients. By simply averaging the percentage change of all the four coefficients, we may conclude that by adjusting for the covariates $\bm X$,  the association between {\it well-being} and {\it anxiety} has been reduced by 34.5\%. However, this reduction is dramatically different from -9.3\% as shown in Panel B, where {\it anxiety} $Y_{A}$ is treated as the response and {\it well-being} as a predictor in the regression model. The observation here is that the effect size is heavily dependent on which outcome variable is used as the ``response'' or ``predictor'' in regression models. 
But ideally, a measure of partial/marginal associations and their difference should be invariant, and not depend on the choice of the ``response'' or ``predictor'' in regression models. Such a measure is particularly needed when the cause/effect relationship (e.g., between {\it well-being} and {\it anxiety}) is bi-directional rather than one-directional.

The regression coefficient method is also sensitive to the choice of model even if the ``response'' variable remains the same. Panel C of Table \ref{tab:ex1} shows when a stereotype model \citep{anderson1984regression,fernandez2019method} is used to fit {\it anxiety}, the change in the regression coefficient is -4.6\% as compared to -9.3\% in Panel B. Now, the question is: which result do we report as the size of the  moderation effect, 4.6\%, 9.3\%, or 34.5\%? The conclusions are contradictory to each other considering that a change of 10\% is the rule of thumb to determine a clinical significance. 


To assess partial association between mixed data and therefore evaluate moderation effects of other risk factors, we develop a unifying framework using the surrogate method. The new framework includes an effect size measure on the $[-1, 1]$ scale that is (i) invariant to the choice of response, (ii) robust to the choice of regression model, and (iii) analogous to Pearson correlation in terms of having similar interpretation. Our approach borrows and advances the recently developed {\it surrogate idea}, which has been approved useful for tackling the discreteness of data \citep{liu2018residuals, cheng2021surrogate, liu2021assessing, liu2022new,  greenwell2018residuals, li2021passo}. Specifically for mixed data analysis, we propose to {\it map the residual randomness (after adjusting for covariates) to the same continuous scale, regardless of the outcome variable being binary, ordinal or continuous}. This is achieved by defining a so-called unified residual that broadly applies to discrete/continuous outcomes and a wide range of regression models. Using this residual, we are able to show that studying partial association between mixed-type outcomes is equivalent to studying marginal association between their residuals. This result allows us to develop an effect size measure that generalizes Kendall's tau \citep{kendall1938new}. The generalization is in the sense that (i) our measure gauges the size and reveals the direction of partial association after adjusting for covariates, yet it reduces to Kendall's tau in the absence of covariates; and (ii) it is exactly equal to the tau-based measure in \citet{liu2021assessing} for cumulative link models, yet it applies more broadly to general regression models and it remains valid for mixed data. 


Compared to \citet{zhu2012nonparametric} and \citet{jiang2014identifying} which studied hypothesis testing for mixed data, our framework offers a [-1,1]-scale measure, a partial regression plot, and a testing procedure that applies to both simple and composite null hypotheses. As demonstrated in \citet{liu2021assessing}, it is desirable to establish a single framework that can yield measures, graphics, and $p$-values altogether for studying partial association. These different inferential tools can complement and strengthen each other so as to achieve a fuller inference. Compared to \citet{liu2021assessing} which focused on ordinal data and cumulative link models, our methodological development has a much broader scope. It advances the theory of the surrogate method \citep{liu2018residuals, cheng2021surrogate, liu2021assessing, liu2022new} by showing that surrogate residuals can be generated without using a latent structure. As a result, it enables us to study different types of data in the same framework. It also offers the flexibility of permitting any commonly used parametric models for covariate adjustments, which can accommodate different  conventions or preferences in different domains. Without these methodological advancements, statistical analysis of YOU Survey and evaluation of the COVID-19 impact could be cumbersome, if not possible, and it might not have generated statistical results and domain insights as detailed in Section 4.

The rest of this article is organized as follows. We introduce in Section \ref{sec:data} the YOU Survey dataset, and in Section \ref{sec:method} the proposed framework for assessing partial association for mixed data. A more detailed and in-depth analysis of our well-being data is presented in Section \ref{sec:realdata} with discussions of our empirical findings. In Section \ref{sec:simu}, we mimic the setting of our well-being study and carry out simulation studies to further demonstrate the advantages of our framework. The paper is concluded by a discussion in Section \ref{sec:summary}.


\section{YOU Survey Data}\label{sec:data}

\begin{figure}
	\centering
	\includegraphics[width=0.8\textwidth]{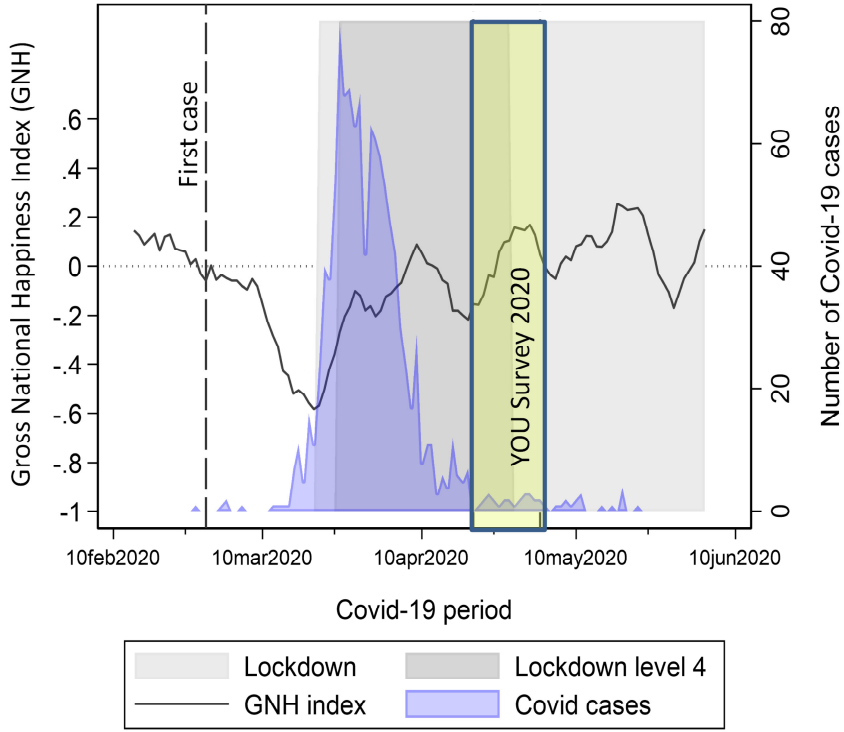}
	\caption{Timeline for the passage of COVID-19 in New Zealand, 2020 (adapted from Figure 1 in \citealp{morrison2021impact}).}
	\label{fig:timeline}
\end{figure}

The YOU Survey data were collected through questionnaires completed by the first-year college students from a major university in New Zealand.\footnote{The data come from responses to the YOU Student Well-being Survey questionnaire delivered on-line via Qualtrics. The data have been stripped of any identifying information and only includes responses which students have agreed to release for the purpose of the research.  The project has received approval from the university's Human Ethics Committee (approval number \#: 0000027270 YOU - Student Wellbeing Study).} 
The YOU project started in the year of 2019, when no one could foresee the once-in-a-lifetime pandemic would break out a year later. 
The intention was to systematically monitor and study the general well-being of college students so as to facilitate policy making. 	
In our study, responses to an identical on-line YOU Survey were collected respectively from new undergraduates in April/May 2019 and again in April/May 2020, a month after the first case of COVID-19 had been detected in the country (see the timeline in Figure~\ref{fig:timeline}). The responses from the 2019 in-take of otherwise similar students constituted a counterfactual against which we were able to compare the well-being responses of those who lived through the pandemic in 2020. 
Moreover, neither the pre- or early pandemic YOU survey carry any mention of COVID-19.  In both surveys, students were asked to answer a suite of well-being questions without being asked to focus on the pandemic. 

The questionnaire in the survey included two sets of questions. 
The first set used four different instruments to measure students' well-being and mental health (i.e., the World Health Organization Well-being Index (WHO-5) \citep{topp20155}; the Generalized Anxiety Disorder Scale (GAD7) \citep{toussaint2020sensitivity,spitzer2006brief}; psychological distress (PHQ-9) \citep{wittkampf2007diagnostic}; and life satisfaction (SWLS) \citep{diener1985satisfaction}). The answers to these questions were collected and serve in our study as four outcome variables: {\it well-being}, {\it anxiety}, {\it depression}, and {\it satisfaction}. The use of different instruments leads to mixed types of outcomes. For example, {\it well-being} and {\it depression} are recorded using numerical scores from 0 to 100, but {\it anxiety} and {\it satisfaction} only have five ordered categories.  As elaborated in the introduction, it is the mixed nature of different types of outcomes that presents technical challenges in the analysis of well-being and mental health data. 	

Another set of questions were designed to capture students' concerns in their college life. Concerns that may impact well-being and mental health include {\it financial strain}, {\it physical healthiness}, {\it loneliness}, and {\it accommodation}. In addition, the data also include demographic variables, {\it age} and {\it gender}, which are controlled in our association and moderation analysis as they are known to have non-negligible influence on well-being and mental health. Brief variable descriptions can be found in Table~S2 in the supplemental materials. Table \ref{tab:datasum} shows summary statistics of numeric variables for each cohort, for which the number of observations is 1224 and 1209, respectively. 

\begin{table}
	\fontsize{11}{12}\selectfont 
	\centering
	\caption{Summary statistics for numerical variables stratified by cohort} \label{tab:datasum}
	\resizebox{\textwidth}{!}{
		\begin{tabular}{lrrrrrrrrrrr}
			\toprule
			& \multicolumn{5}{c}{2019} && \multicolumn{5}{c}{2020}\\
			\cline{2-6} \cline{8-12}
			Variable & Min & Med. & Max & Mean & Std. & & Min & Med. & Max & Mean & Std. \\
			\midrule
			Wellbeing & 8 & 56 & 100 & 53.25 & 17.03 && 4 & 52 & 100 & 51.67 & 18.24 \\ 
			Anxiety & 1 & 3 & 5 & 3.34 & 1.03 && 1 & 3 & 5 & 3.22 & 1.04 \\ 
			Depression & 0 & 29 & 100 & 32.91 & 21.40 && 0 & 29 & 100 & 32.20 & 20.28 \\
			Satisfaction & 1 & 6 & 7 & 5.11 & 1.47 && 1 & 6 & 7 & 5.11 & 1.44 \\	
			Financial strain & 1 & 2 & 5 & 2.50 & 1.15 && 1 & 2 & 5 & 2.26 & 1.09 \\ 
			Healthiness & 1 & 4 & 5 & 3.48 & 0.84 && 1 & 4 & 5 & 3.67 & 0.82 \\
			Loneliness & 1 & 2 & 5 & 2.54 & 1.05 && 1 & 2 & 5 & 2.25 & 1.05 \\ 
			Age & 15 & 18 & 35 & 18.73 & 2.29 && 15 & 18 & 35 & 18.91 & 2.64 \\
			\bottomrule
		\end{tabular}
	}
\end{table}

This design feature of the YOU survey has helped avoid the possibility of the ``focusing illusion'' (\citealp{schkade1998does}, \citealp{kahneman2011thinking} p. 402-406) -- the tendency to exaggerate the effects of the single event being focused on. Our survey differs in this important respect from those studies (see Supplementary Materials) which explicitly ask students to focus on the effect of the pandemic on their well-being. In addition, unlike many existing works in which the survey had to rely on the student’s recall of pre-COVID-19 conditions rather than being able to use measurements taken at the time, the YOU survey data reduces such a ``recall bias'' \citep{kopec1990bias}.   
The YOU survey was set up simply to monitor student well-being over time. It was not intended to study the impact of COVID-19 but serendipitously turns out to provide a more reliable research design for assessing the well-being effects of the pandemic.

Our data show that there is no clinically meaningful difference between the two cohorts in the mean statistics measuring well-being and mental health. Part of the reason may have to do with the relatively low penetration of the virus in New Zealand as well as the rapid response of the New Zealand government and tertiary institutions both of which contributed to the short-lived well-being response observed nationally \citep{morrison2021impact}. However, it may be too quick to conclude that COVID-19 had little impact on the well-being and mental health of first year students in New Zealand. In fact, our case study shows that a simple comparison of summary statistics is misleading, and a more subtle response would have been concealed without a deeper analysis of the data. Using the proposed method, we find that the arrival of the pandemic has changed the underlying association between student well-being and anxiety, another important measure of mental health status.



\section{Methodology}\label{sec:method}

\subsection{A unified residual}

We study the pairwise partial association between $K$ outcome variables $Y_1,...,Y_K$, which are allowed to be measured on heterogeneous scales (i.e., continuous, binary, and ordinal). To adjust for a given set of covariates $\bm X$, our approach considers general regression models $Y_k \sim \bm X$ and captures the residual randomness using a {\it unified residual} $R_k$. We show that studying the partial association between $Y_k$ and $Y_j$ is equivalent to studying the marginal association between $R_k$ and $R_j$. The key idea is to express the residual randomness (after covariate adjustments) in the same continuous scale even when outcome variables are measured in different scales. This idea is implemented with the aid of the residual introduced below.  

For outcome $Y$, we consider a general regression model for adjusting for covariates $\bm X$
\begin{align}
	Y\mid (\bm X = \bm x) \sim F(y; \bm x, \bm \beta), \label{eq:model1}
\end{align}
where $F(\cdot)$ is the cumulative distribution function (CDF) determined by a chosen regression model and $\bm \beta$ the associated regression coefficients. Model \eqref{eq:model1} is general enough to cover virtually all commonly used regression models including, for instance, linear models for continuous outcomes ($F(\cdot)=\Phi(\cdot)$) and generalized linear models for binary or ordinal outcomes \citep{mccullagh1989generalized}. Unlike \citet{liu2021assessing}, the generalized linear models considered in this paper are not limited to cumulative link models (e.g., the proportional odds model), but include the adjacent category logit model and ordered stereotype model as well \citep{liu2005analysis,agresti2010analysis,tutz2020ordinal,kosmidis2021mean}. 
Upon an appropriate choice of Model \eqref{eq:model1} for the outcome variable $Y$, a unified residual can be defined as follows.

\begin{definition} \label{def1}
	If Model \eqref{eq:model1} specifies the distribution of $Y$, we define a  residual variable $R$ as
	\begin{align}
		R(Y=y \mid \bm x, \bm \beta)=S(Y=y \mid \bm x, \bm \beta)-\mathbb{E}(S \mid \bm x, \bm \beta). \label{eq:sr1}
	\end{align}
	Here for any realized value $y$,
	\begin{align}
		S(y; \bm x, \bm \beta) \sim U(F(y_{-}; \bm x, \bm \beta), F(y; \bm x, \bm \beta)), \label{eq:varS}
	\end{align}
	and $F(y_{-}; \bm x, \bm \beta)=\lim_{z\to y^{-}}F(z; \bm x, \bm \beta)$, where $\lim_{z\to y^{-}}$ is the one-sided limit as $z$ approaches $y$ from the left.
\end{definition}

To better understand this unified residual, we start by considering a continuous outcome $Y$ with a linear regression model, where the error term follows $N(0, \sigma^2)$. In this case, 
\begin{align*}
	s(y; \bm x, \bm \beta)=\Phi(r/\sigma), \quad \text{and} \quad r(y; \bm x, \bm \beta)=\Phi(r/\sigma)-0.5. 
\end{align*}
Here, $r=y-\mathbb{E}(Y|\bm x)$ is the classical residual and $\Phi(\cdot)$ is the CDF of the standard normal distribution. Our definition maps the classical residual $r$ to a scale of (-1/2, 1/2), and it reduces to the Cox-Snell residual \citep{cox1968general}. Throughout the paper, we use an upper-case letter (e.g., $R$ or $S$) to denote a random variable and a lower-case letter (e.g., $r$ or $s$) an observation.

If the outcome $Y$ is ordinal (i.e., $Y \in \{1, 2,..., J\}$), the $S$ variable in Definition~\ref{def1}
\begin{align*}
	S(y; \bm x, \bm \beta) \sim U(F(y-1; \bm x, \bm \beta), F(y; \bm x, \bm \beta)), \quad y=1,2,...,J. 
\end{align*}
As a special case, if the outcome $Y$ is binary, e.g., $Y \in \{0, 1\}$, the formula is simply 
\begin{align*}
	\begin{cases}
		S(y; \bm x, \bm \beta) \sim U(0, F(0; \bm x, \bm \beta)) & \text{ if } Y=0;\\
		S(y; \bm x, \bm \beta) \sim U(F(0; \bm x, \bm \beta), 1) & \text{ if } Y=1.
	\end{cases} 
\end{align*}
In fact, the $S$ variable is the {\it general surrogate variable} proposed in \cite{liu2018residuals} and \cite{liu2021assessing} (see Section 7 of the two papers), where the core idea is to use a continuous variable $S$ as a surrogate of the original ordinal variable to draw inference. However, \cite{liu2018residuals} and \cite{liu2021assessing} did not investigate further the utility of $S$ in statistical inference. This paper shows that given mixed outcomes, the variable $S$ defined in (\ref{eq:varS}) enables us to map the residual randomness (after adjusting for covariate) to the same continuous scale, regardless of $Y$ being continuous, binary, or ordinal. This result is stated in Property~\ref{property:R-dist}. 

\begin{property}\label{property:R-dist}
	If Model \eqref{eq:model1} specifies the distribution of $Y$, then the residual $R$ defined in (\ref{eq:sr1}) follows the uniform distribution on (-1/2, 1/2).
\end{property}

Recall that if both outcome variables $Y_1$ and $Y_2$ were continuous, their partial association could be studied through their residuals obtained from linear models $Y_{1} \sim \bm X$ and $Y_{2} \sim \bm X$ \citep{fisher1924distribution,johnson2007applied}. Property~\ref{property:R-dist} makes this residual-based partial association analysis applicable when the outcomes are recorded on different scales. We will justify in the next subsection that to examine partial association between mixed outcomes, it suffices to study marginal association between their residuals defined in (\ref{eq:sr1}). In what follows, we present another two properties of the residual. 

\begin{property}
	The residual $R$ defined in (\ref{eq:sr1}) is invariant to the following transformations of the outcome variable $Y$: (1) any linear transformation if $Y$ is continuous; and (2) any monotonic transformation if $Y$ is ordinal or binary.  
\end{property}

Property 2 follows straightforwardly if $Y$ is binary or ordinal as a monotonic transformation does not change the order of the categories. If $Y$ is continuous, a linear transformation $Y^*=aY+b$ results in a residual $r^*(y; \bm x, \bm \beta)=\Phi((a \cdot r)/(a\cdot\sigma))-0.5 =\Phi(r/\sigma)-0.5 = r(y; \bm x, \bm \beta)$. The invariant to linear/monotonic transformations implies that our residual-based partial association analysis will be robust in the sense that it does not depend on the range of continuous outcomes or the label of categorical outcomes.

\begin{property}
	\label{propty:specialcase}
	If Model~(\ref{eq:model1}) is a cumulative link model with the link function being $G(\cdot)$, \citep{liu2018residuals}'s surrogate residual $R_{LZ}$ is equivalent to the residual $R$ in the sense that a transformation of $R_{LZ}$, i.e., $G(R_{LZ})-1/2$, follows the same distribution of $R$. This statement is true both unconditionally and conditionally on $(Y=y, \bm X=\bm x)$.
\end{property}

For the class of cumulative link models for discrete data, Property~\ref{propty:specialcase} says that Liu-Zhang's surrogate residual, which is generated using the latent model structure, is in fact equivalent to the residual in (\ref{eq:sr1}), which nevertheless does not rely on the notion of latent variables. Note that the simulation of the continuous surrogate variable $S$ in (\ref{eq:varS}) is simply a random draw from a uniform distribution. The lower and upper bounds are determined by the data and Model (\ref{eq:model1}), but other than that, the generating process of $S$ does not re-use any model assumptions. With this said, Definition~\ref{def1} broadens the scope of the surrogate method used in \citep{liu2018residuals, cheng2021surrogate, liu2021assessing, liu2022new} in the sense that  it applies to regression models as general as in Model~(\ref{eq:model1}), which encompasses almost all parametric models.


\begin{remark}
	For discrete components of mixed outcomes, the generation of the general surrogate variable $S$ in (\ref{eq:varS}) is a simulation that seems to have added ``random noise'' to discrete data. We show in Section 2.3 a surprising result; that is, our rank-based effect size measure is not influenced by the added randomness incurred in the simulation of $S$.
\end{remark}

\begin{remark}
	When an outcome is discrete, simulating continuous data for better visualization purposes is not new. See the randomized quantile residual \citep{dunn1996randomized} and a more general jittering technique \citep{chambers2018graphical}. The current work is among a systematic development to show that the simulated continuous outcome $S$ can be used as a ``surrogate'' to solve a broader scope of inference challenges, which include how to establish a coherent inference framework to study partial association between  mixed outcomes. 
\end{remark}

	\subsection{General $\mathcal{T}$ as an association measure}

In this subsection, we propose a new association measure $\mathcal{T}$. We show that it can gauge the size and capture the direction of partial association as well as marginal association between mixed outcomes. The results below lay the foundation for defining the  measure $\mathcal{T}$.

\begin{theorem}
	\label{thm:equiv}
	Suppose that two outcome variables $Y_1$ and $Y_2$ follow Model~(\ref{eq:model1}), i.e., $Y_1 \sim F_{Y_1|\bm X=\bm x}(y; \bm x, \bm \beta_1)$ and $Y_2 \sim F_{Y_2|\bm X = \bm x}(y; \bm x, \bm \beta_2)$. Let $R_1$ and $R_2$ be the corresponding residual variables as defined in Definition 1. Then, for any given value $\bm x$, we have
	\begin{align}
		(Y_1 \perp \!\!\! \perp Y_2)|(\bm X = \bm x) \Leftrightarrow (R_1 \perp \!\!\! \perp R_2)| (\bm X = \bm x). \label{eq:thm1}
	\end{align}
	Furthermore, we have
	\begin{align}
		R_1 \not\!\perp\!\!\!\perp R_2  \Rightarrow (Y_1 \not\!\perp\!\!\!\perp Y_2)| (\bm X =\bm x)\ \text{for some}\ \bm x. \label{eq:thm12}
	\end{align}
\end{theorem}

Theorem \ref{thm:equiv} provides the key result underlying our framework for analysis of partial association between mixed outcomes. The result in \eqref{eq:thm1} provides a sufficient and necessary condition for the conditional independence between the outcome variables $Y_1$ and $Y_2$. It says we only need to check the conditional independence between the residual variables  $R_1$ and $R_2$. The result in \eqref{eq:thm12} implies that analyzing partial association between $Y_1$ and $Y_2$ can be reduced to analyzing the marginal association between the residuals $R_1$ and $R_2$. In particular, if $R_1$ and $R_2$ are shown to be correlated using any of our proposed measure, graphic, or $p$-value, we can conclude that $Y_1$ and $Y_2$ are dependent for some values of $\bm X$.

Based on Theorem \ref{thm:equiv}, we define a general partial association measure
\begin{align}
	\mathcal{T}(Y_1, Y_2 : \bm X)=\tau(R_1, R_2). \label{eq:tau}
\end{align}
Here, $\tau(\cdot, \cdot)$ is Kendall's tau \citep{kendall1938new}, which is a rank-based correlation measure defined as 
\begin{align*}
	\tau(R_1, R_2)=P\{(R_1-\check{R}_1)(R_2-\check{R}_2)>0\}-P\{(R_1-\check{R}_1)(R_2-\check{R}_2)<0\},
\end{align*}
where $(\check{R}_1, \check{R}_2)$ are independent copies of $(R_1, R_2)$. Given $n$ pairs of realizations $(\bm r_1, \bm r_2)=\{(r_{1i}, r_{2i})\}_{i=1}^n$, an empirical estimate of our proposed measure $\mathcal{T}$ is 
\begin{align}
	\hat{\mathcal{T}}=\hat{\tau}(\bm r_1, \bm r_2)=
	\begin{pmatrix} n\\ 2 \end{pmatrix}^{-1} \sum_{i<j} \text{sgn}(r_{1i}-r_{1j})\text{sgn}(r_{2i}-r_{2j}), \label{eq:tauhat}
\end{align}
where the sign function $\text{sgn}(u)=1$ if $u>0$, $\text{sgn}(u)=-1$ if $u<0$, and $\text{sgn}(u)=0$ if $u=0$. Without confusion, we use $\mathcal{T}$ in place of $\mathcal{T}(Y_1, Y_2 : \bm X)$ in the rest of the paper.

Similar to Kendall's tau, the range of the general $\mathcal{T}$-measure is $[-1,1]$. It is the difference between the proportions of concordance pairs and the proportion of discordance pairs of the observed residuals $\{(r_{1i}, r_{2i})\}_{i=1}^n$. Concordance (or discordance) refers to situation where the movement from $r_{1i}$ to $ r_{1j}$ and that from $r_{2i}$ to $r_{2j}$ are in the same (or opposite) direction. Therefore, the $\mathcal{T}$-measure reflects the direction  and strength of a monotonic association between the two residual variables $R_1$ and $R_2$. As a monotonic association may not necessarily be linear, the $\mathcal{T}$-measure is potentially more powerful in capturing partial association between $Y_1$ and $Y_2$. In the following section, we will present its properties, which justify (i) the use of the rank-based formula in the $\mathcal{T}$-measure; and (ii) the generality of the $\mathcal{T}$-measure.

\subsection{Properties of the $\mathcal{T}$-measure}
Our first proposition below shows our $\mathcal{T}$ is a general rank-based measure including Kendall's tau as a special case when no covariate needs to be adjusted for.

\begin{proposition}
	\label{prop:nox}
	In the absence of covariates, the $\mathcal{T}$-measure is exactly the same as Kendall's tau, i.e., $\mathcal{T}(Y_1, Y_2)=\tau(Y_1, Y_2)$. 
\end{proposition}
Proposition~\ref{prop:nox} justifies the use of the $\mathcal{T}$-measure for sizing both marginal and partial associations between mixed outcomes. The difference between the two effect sizes 
\[\mathcal{M}_{Y_1Y_2}(\bm X) = \mathcal{T}(Y_1, Y_2 : \bm X) - \mathcal{T}(Y_1, Y_2)\]
reflects the change of association direction/strength before and after an  adjustment of covariates. In our well-being study, this numerical difference is used to evaluate the moderation effects of several risk factors of interest. Our empirical analysis shows that the moderation effect is significantly greater after the college students have experienced epidemic and associated lock-downs. In our analysis, neither of $Y_1$ nor $Y_2$ is treated as a response and the other as a predictor. 

\begin{remark}
	Proposition~\ref{prop:nox} does not hold if a moment-based correlation measure, such as Pearson correlation $\rho(\cdot, \cdot)$, is applied to $R_1$ and $R_2$ (see the measure $\phi_\rho$ in \citealp{liu2021assessing}). The rank-based formula in (\ref{eq:tauhat}) allows the randomness (or ``noise'') added in the simulation of $S_1$ and $S_2$ cancel out when the correlation between $(R_1, R_2)$ is calculated. This is part of the reason we use the rank-based measure in (\ref{eq:tau}). The next proposition further investigate the impact of the added randomness in the simulation of $S$. 
\end{remark}

A similar result to Proposition~\ref{prop:nox} is proved in \citet{liu2021assessing} for ordinal outcomes and cumulative link models. Our result holds for mixed data and much more general models. The next proposition shows \citet{liu2021assessing}'s measure is a special case of the general $\mathcal{T}$-measure.

\begin{proposition}
	\label{prop:specialcase}
	If both $Y_1$ and $Y_2$ are ordinal variables and Model \eqref{eq:model1} is in the class of cumulative link models, the $\mathcal{T}$-measure is exactly the same as \citet{liu2021assessing}'s $\phi_\tau$ i.e., $\mathcal{T}(Y_1, Y_2 : \bm X) = \phi_\tau(Y_1, Y_2 : \bm X)$.
\end{proposition}

\citet{liu2021assessing}'s $\phi_\tau$ for ordinal outcomes requires simulation of surrogate variables. Their simulation uses the latent structure of cumulative link models, and it is different from our simulation from the uniform distribution in (\ref{eq:varS}). Propositions~\ref{prop:specialcase} delivers a surprising result, which basically says that when the assumption of cumulative links holds, the two different simulation processes lead to exactly the same population parameter of partial association. This result implies that the use of different ways to add ``noise'' does not alter our defined association measure. The result also implies that our $\mathcal{T}$-measure subsumes \citet{liu2021assessing}'s $\phi_\tau$, and it applies more broadly without relying on cumulative link models for covariate adjustments.

To summarize the results in Propositions~\ref{prop:nox} and \ref{prop:specialcase}, our $\mathcal{T}$-measure is general enough to (i) measure both marginal and partial associations; (ii) apply to mixed outcomes; and (iii) allow the use of general regression models for covariate adjustments. The last proposition presents an invariant property of the $\mathcal{T}$-measure.
\begin{proposition}
	\label{prop:invariant}
	The $\mathcal{T}$-measure  is invariant to monotonic transformations of either or both residual variables $R_k$'s in (\ref{eq:tau}), i.e., $\mathcal{T}(Y_1, Y_2 : \bm X)=\tau(R_1, R_2)=\tau(h_1(R_1), h_2(R_2))$ where $h_k(\cdot)$'s are monotonic transformation functions.
\end{proposition}

\begin{figure}
	\centering 		
	\includegraphics[scale=0.6]{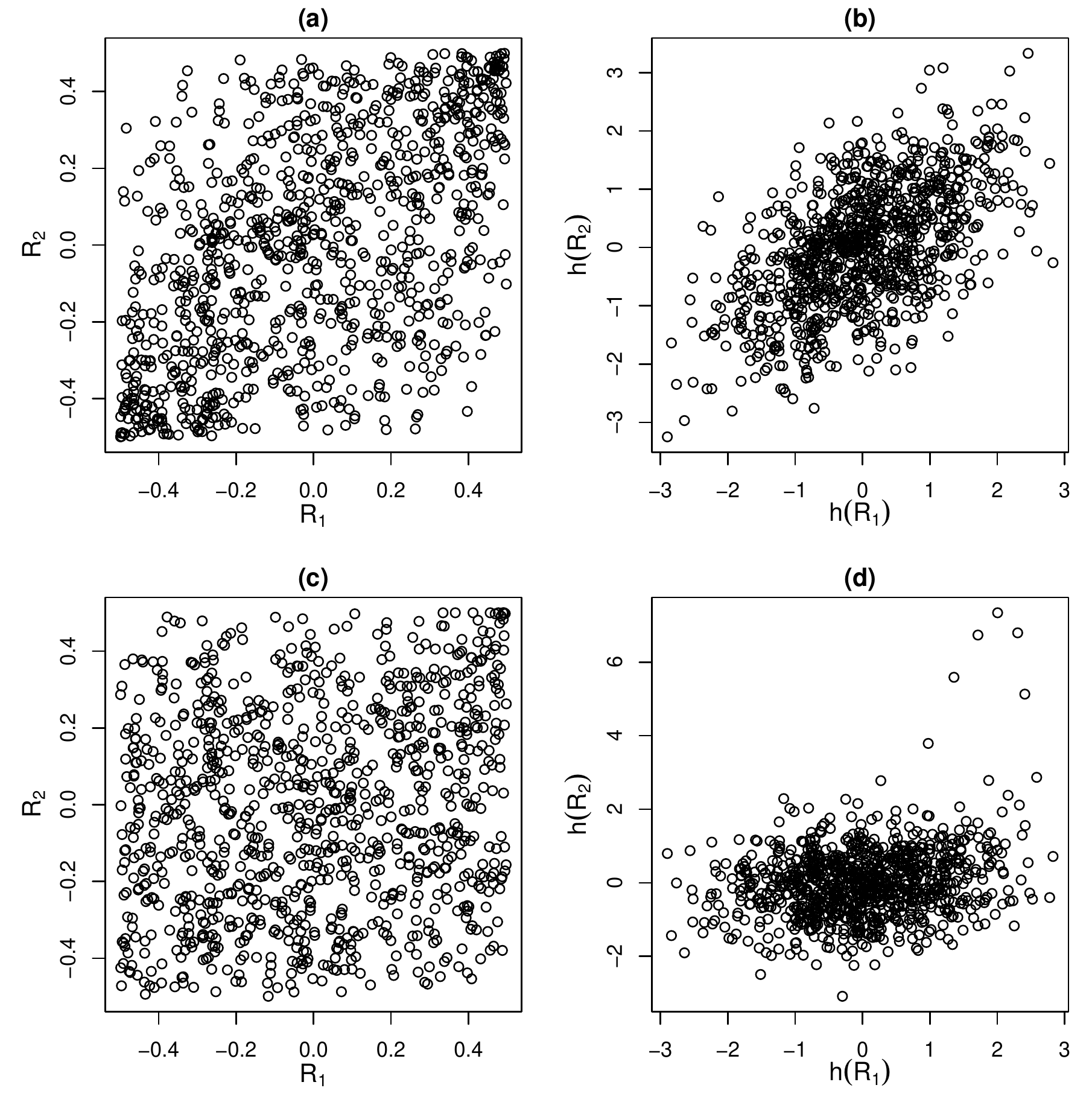}
	\caption{Visualizing partial association using partial regression plots. The left plot scatters $R_1$ versus $R_2$, while the right plot scatters $h(R_1)$ versus $h(R_2)$.} \label{fig:ex1}
\end{figure}

Proposition \ref{prop:invariant} is a direct result from the definition of Kendall's tau. Based on this result, we propose to apply the transformation $h_1(c)=h_2(c)=h(c)=\Phi^{-1}(c+1/2)$ to the residuals to better visualize partial association. The benefit can be seen in Figure \ref{fig:ex1}(a)-(b) where (a) scatters $R_1$ versus $R_2$ while (b) scatters $h(R_1)$ versus $h(R_2)$. Such plots are called partial regression plots in the literature. While we can hardly see any associations in Figure \ref{fig:ex1}(a), a moderately strong association  can be easily spotted in Figure \ref{fig:ex1}(b) with the transformation $h(\cdot)$ applied. Another benefit can be seen in the second row of Figure~\ref{fig:ex1} where the strength and shape of partial association are different from those in the first row. While the difference between Figure~\ref{fig:ex1}(c) and (a) is subtle, 
Figure~\ref{fig:ex1}(d) manifests a clear distinction of the association structure from that in (b). Moreover, Figure~\ref{fig:ex1}(d) shows the potential of our residual to unveil nonlinear partial association, which will be demonstrated further in the analysis of simulated and real data sets. In view of these benefits, we recommend using the transformation for visualization given that it does not alter the size of the inspected partial association.

\subsection{Inference of the $\mathcal{T}$-measure}

So far, we have studied the $\mathcal{T}$-measure itself. To draw inference about this measure, we need to simulate $s_i$ from (\ref{eq:varS}) which relies on estimated probabilities $\hat{F}(y_i; \bm x_i, \hat{\bm \beta})$. Our first result shows that $\hat{\mathcal{T}}$ in \eqref{eq:tauhat} is a consistent estimate of $\mathcal{T}$ under certain conditions. 

\begin{theorem}
	\label{thm:consist}
	The estimate $\hat{\mathcal{T}}$ in \eqref{eq:tauhat} is consistent, i.e., $\hat{\mathcal{T}}-\mathcal{T}=o_p(1)$, if the following two conditions hold: (i) the distribution function $F(\cdot)$ in \eqref{eq:model1} is continuous in model parameter vector $\bm \beta \in \mathbb{R}^d$ for any given observed data point $(y, \bm x)$; and (ii) the regression coefficient estimate $\hat{\bm \beta}_k$ is consistent, i.e., $\hat{\bm \beta}_k-\bm \beta_k=o_p(1)$, for $k=1,2$.
\end{theorem}

If one of the outcome variables is discrete, the simulation of $s_i$ introduces an additional layer of randomness, which will be passed to the estimate in \eqref{eq:tauhat}. To reduce variability and stabilize the estimate, we follow the recommendations in (\citealp{hong2010prediction} and \citealp{liu2021assessing}) and 
take an average of \eqref{eq:tauhat} based on multiple (e.g., $M$) simulations of $S$. Specifically, 
\begin{align}
	\hat{\mathcal{T}}_M=\frac{1}{M}\sum_{m=1}^{M}\hat{\mathcal{T}}^{(m)}, \label{eq:phihatM}
\end{align}
where $\hat{\mathcal{T}}^{(m)}$ is an estimate in \eqref{eq:tauhat} using the $m$-th simulation of $S$.
In our numerical studies, we use a sufficiently large $M=30$ and report \eqref{eq:phihatM} for inference.

To assess the variability of $\hat{\mathcal{T}}_M$, we follow \citet{liu2021assessing} and use the bootstrap method \citep{efron1994introduction}. It allows us to derive the standard error of $\hat{\mathcal{T}}_M$, confidence intervals for $\mathcal{T}$, and a $p$-value for testing $H_0$: $\mathcal{T}=0$ altogether. Specifically, in the $b$-th bootstrap, we (i) draw a random sample $(\bm y_1^*, \bm y_2^*, \bm x^*)^{(b)}$ from the original sample; (ii) refit Model \eqref{eq:model1} to the bootstrap sample and obtain bootstrap residuals $(\bm r_1^*, \bm r_2^*)^{(b)}$; and (iii) 
compute $\hat{\mathcal{T}}_M^{*(b)}$ based on \eqref{eq:phihatM}. Repeating the steps in (i)-(iii) $B$ times, we have a set of bootstrap estimates $\{\hat{\mathcal{T}}_M^{*(1)},\ldots,\hat{\mathcal{T}}_M^{*(B)}\}$, which gives a bootstrap distribution of $\hat{\mathcal{T}}_M$, denoted by $\hat{F}^*_B(\mathcal{T})$. The standard deviation of $\hat{F}^*_B(\mathcal{T})$ can be used as an estimate of the standard error of $\hat{\mathcal{T}}_M$. A two-sided $100(1-\alpha)\%$ confidence interval for $\mathcal{T}$ can be constructed as $(\hat{F}^{*-1}_B(\alpha/2), \hat{F}^{*-1}_B(1-\alpha/2))$. To test the simple null hypothesis $H_0$: $\mathcal{T}=0$ versus $H_1$: $\mathcal{T}\ne 0$, a $p$-value can be determined by $2\min\{\hat{F}^*_B(0), 1-\hat{F}^*_B(0)\}$. Additionally, the bootstrap method provides extra flexibility that allows testing a composite hypothesis such as $H_0: |\mathcal{T}|< \delta$ where $\delta$ is a pre-specified association strength level. In this case, the $p$-value is $2\min\{\hat{F}^*_B(\delta), \hat{F}^*_B(-\delta)\}$. 


	\section{Analysis of the YOU Survey data}\label{sec:realdata}

Following the motivating analysis in Section \ref{sec:intro}, we carry out an in-depth analysis of the data for both pre-pandemic and early pandemic periods, and discuss the
impact of COVID-19 on the well-being of college students and its implications for university policy makers and healthcare providers.

\subsection{Moderation effects of risk factors based on partial association}

First of all, we evaluate whether or not and to what extent risk factors moderate the correlation between the four outcome variables {\it well-being}, {\it anxiety}, {\it depression}, and {\it satisfaction}. The risk factors considered include {\it financial strain}, {\it physical healthiness}, {\it loneliness}, {\it accommodation}, {\it age}, and {\it gender}. Table \ref{tab:realdata1} displays the association analysis result without and with the adjustment of risk factors (i.e., marginal versus partial associations). We observe that (i) all the partial associations maintain the same direction as the corresponding marginal associations; and (ii) the strengths of all the pair-wise associations have been reduced after adjusting for risk factors. These observations suggest that the risk factors contribute to the correlations between outcomes variables. To quantitatively evaluate the size of such moderation effects, we calculate the percentage change of the association. The result is summarized in  Table \ref{tab:realdata2}. The sizes of association reduction are significant statistically for all the pairs. The lowest reduction  is $-18.9\%$ and the highest is $-39.3\%$. This result implies that the 6 risk factors considered here moderate the relationship between well-being and mental health measures in a significant way. But the investigation so far has not evaluated the impact of each single moderator. In the next section, we carry out moderator-specific analysis and monitor its effect change before and after the COVID-19 outbreak. 

\begin{table}
	\centering
	\caption{Marginal and partial associations between the four outcomes {\it well-being},  {\it anxiety}, {\it depression} and {\it satisfaction} for the 2020 student cohort.  Shown are the estimates of our measure $\mathcal{T}$ and their standard errors (in the parenthesis). } \label{tab:realdata1}
	\vspace{0.2cm}
	\begin{tabular*}{\textwidth}{l@{\extracolsep{\fill}}*{6}{r}}
		\toprule
		& \multicolumn{3}{c}{Marginal association} & \multicolumn{3}{c}{Partial association}\\
		\cline{2-4} \cline{5-7}
		& {\it anxiety} &{\it depression}& {\it satisfaction } & {\it anxiety} &{\it depression}& {\it satisfaction }\\
		\midrule
		{\it well-being}& -0.374 & -0.472 & 0.466 & -0.247& -0.383 & 0.341 \\
		&  \scriptsize{\textit{(0.019)}} & \scriptsize\textit{(0.015)} & \scriptsize \textit{(0.017)}    & \scriptsize \textit{(0.017)} & \scriptsize \textit{(0.016)} & \scriptsize \textit{(0.017)} \\
		{\it anxiety}   &  & 0.421 & -0.341 &   & 0.294 & -0.207 \\
		&  &   \scriptsize \textit{(0.018)} & \scriptsize \textit{(0.021)} &    & \scriptsize \textit{(0.016)} & \scriptsize \textit{(0.017)} \\
		{\it depression} &  &   & -0.421 &    &  & -0.310\\
		&  &  &   \scriptsize \textit{(0.018)} &   &  & \scriptsize \textit{(0.017)}  \\
		\bottomrule
	\end{tabular*}
\end{table}

\begin{table}[h!] 
	\centering
	\caption{Moderation effects of the risk factors (presented using the percentage change of association after adjusting for risk factors). In the parenthesis are the standard errors.} \label{tab:realdata2} 
	\vspace{0.2cm}
	\begin{tabular}{lccc}
		\toprule
		& {\it anxiety}&  {\it depression} &  {\it satisfaction} \\
		\midrule
		{\it well-being}&   -34.0\% & -18.9\%  & -26.8\% \\
		&   \scriptsize {\it (2.6\%)} & \scriptsize {\it (2.1\%)} & \scriptsize {\it(2.0\%)}\\
		{\it anxiety}  &  & -30.2\% & -39.3\%\\
		& &\scriptsize {\it  (2.2\%)} & \scriptsize {\it(2.7\%)}\\
		{\it depression} &  &  &  -26.4\% \\
		&  & & \scriptsize {\it(2.3\%)}\\
		\bottomrule
	\end{tabular}
\end{table}

\begin{figure}
	\centering 		
	\includegraphics[scale=0.8]{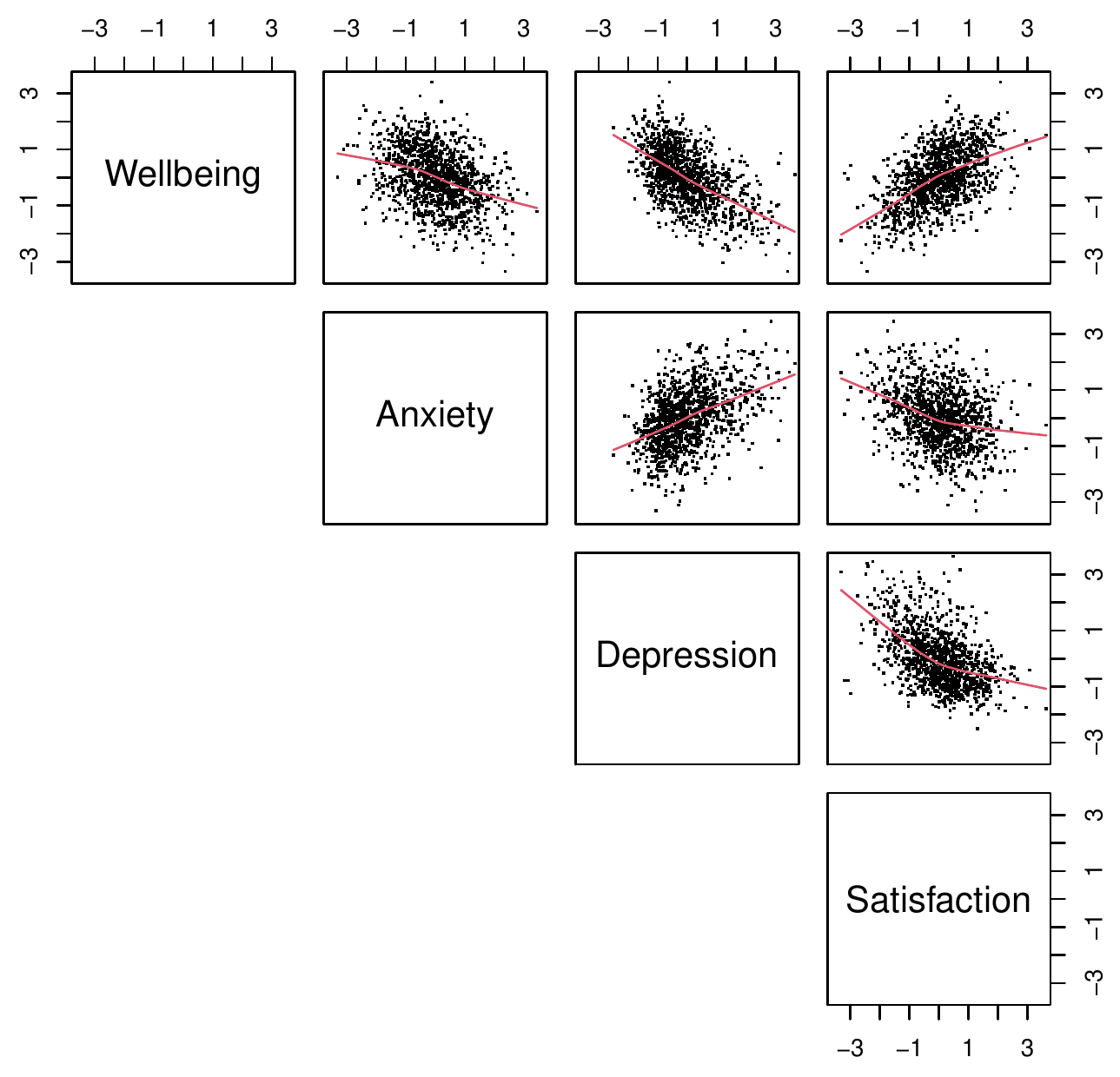}
	\caption{Partial regression plot of each pair of {\it well-being},  {\it anxiety}, {\it depression} and {\it satisfaction}. Red-colored line is the fitted smoothing curve using the LOWESS method.} \label{fig:parplot}
\end{figure}

Our method also enables us to further inspect the structure of the partial associations between {\it well-being},  {\it anxiety}, {\it depression} and {\it satisfaction}. The micro association structures are visualized in Figure \ref{fig:parplot}. We can probably say that all the associations are ``approximately'' linear. However, we notice that the partial association between {\it satisfaction} and other outcomes exhibits some degrees of non-linearity. In particular, this non-linearity may be non-ignorable in the pair of {\it satisfaction} and {\it depression}. 

\subsection{Changes of moderation effects due to COVID-19}

We monitor potential changes of moderation effects before and after the strike of COVID-19. Specifically, we examine {\it financial strain}, {\it physical healthiness}, {\it loneliness}, and {\it accommodation} -- one at a time -- in moderating the correlation between {\it well-being} and {\it anxiety}. While  examining the effect of each individual moderator by $\frac{\mathcal{T}(Y_{W}, Y_{A} : X) - \mathcal{T}(Y_{W}, Y_{A})}{\mathcal{T}(Y_{W}, Y_{A})}$, we always control the effects of {\it age} and {\it gender}. The result is summarized in Table~\ref{tab:realdata3} for the 2019 and 2020 cohort, respectively. It is clear that all the four risk factors show an increased moderation effect in the 2020 cohort. In particular, the increase is significant, both statistically and clinically, for {\it physical healthiness}, {\it loneliness}, and {\it accommodation} (35\%, 35\%, and 24\%). The significant increase indicates that one/two months after the strike of COVID-19, the way students' well-being interacts with anxiety is considerably moderated by their physical health status, feeling of being lonely or isolated from others, and living space and environment (e.g., home with parents, student hall, or renting housing). This finding has several implications. As remote learning has become common, it is important to encourage students to maintain/improve their physical health and stay socially connected with others. The trend of remote learning also calls for re-thinking of how to provide a comfortable learning/living space, either in a student hall or at home. New policies and programs should be developed to help college students cope with the changes brought by COVID-19.

\begin{table}
	\centering
	\caption{Moderation effects of {\it physical healthiness}, {\it loneliness}, and {\it accommodation} and {\it financial strain} in the 2019 and 2020 student cohorts.} \label{tab:realdata3}
	\vspace{0.2cm}
		\begin{tabular}{lcccc}
			\toprule
			& {\it physical healthiness} & {\it loneliness} & {\it accommodation} & {\it financial strain} \\
			\midrule
			2019 cohort &	-0.23 \footnotesize{\textit{(0.02)}} & -0.17 \footnotesize{\textit{(0.01)}} & -0.17 \footnotesize{\textit{(0.01)}} & -0.21 \footnotesize{\textit{(0.01)}}\\
			2020 cohort &	-0.31 \footnotesize{\textit{(0.02)}}	& -0.23 \footnotesize{\textit{(0.02)}} &	-0.21 \footnotesize{\textit{(0.01)}} &-0.23 \footnotesize{\textit{(0.02)}}\\	
			\midrule
			Numerical change  & 0.08 & 0.06 & 0.04 & 0.02 \\
			Percentage change & 35\% & 35\% & 24\% & 10\%\\
			$p$-value & 0.03 & $<0.01$ & $<0.01$ & 0.58 \\
			\bottomrule
		\end{tabular}
\end{table}



\section{Simulation}\label{sec:simu}
\subsection{Setting}
We carry out simulation studies to show our framework allows us to fully study partial association between mixed data. A full study includes quantitative, graphical, and testing assessments. We simulate synthetic well-being data sets by mimicking the data structure of the YOU Survey (see Table \ref{tab:ex1}). Specifically, we simulate a pair of outcome variables {\it well-being} $\tilde{Y}_{W}$ and {\it anxiety} $\tilde{Y}_{A}$ as well as a same set of risk factors $\tilde{\bm X}$. Here, we use $\tilde{.}$ to differentiate the synthetic variables from the real ones. Each risk factor is simulated independently using its empirical distribution. The ordinal variable {\it anxiety} $\tilde{Y}_{A}$ is simulated from an adjacent category logit model $Y_A \sim \bm X$ using the estimated regression coefficients from the real data analysis. Given $\tilde{Y}_{A}$ and $\tilde{\bm X}$, the continuous outcome {\it well-being} $\tilde{Y}_{W}$ is generated using the estimated linear model $Y_W \sim Y_A+\bm X$. The coefficient estimates of these two models can be found in Table~S1 in the supplementary materials. The numerical advantages to be demonstrated include (Section 3.1) the superiority over the traditional regression coefficient method in evaluating the moderation effect of risk factors (i.e., the difference between partial and marginal association measures); (Section 3.2) the effectiveness in visualizing partial association and capturing non-linear associations; and (Section 3.3) the higher power in hypothesis testing as compared to the likelihood ratio test. 

\subsection{Evaluating the moderation effects of risk factors}
As discussed in the introduction, it is of particular interest in well-being studies to monitor how risk factors moderate the association between outcome variables. It reflects, in our case study, the impact of COVID-19 on the well-being status of college students. We have seen in Table \ref{tab:ex1} the inconsistent and confusing result of the traditional regression coefficient method, which requires choosing one outcome as a predictor and the other as the response. In this section, we analyze 1000 synthetic data sets to confirm the inconsistency is a systematic issue. Our analysis also evaluates the effectiveness of our proposed method in moderation analysis. 

\begin{figure}
	\centering 		
	\includegraphics[scale=0.55]{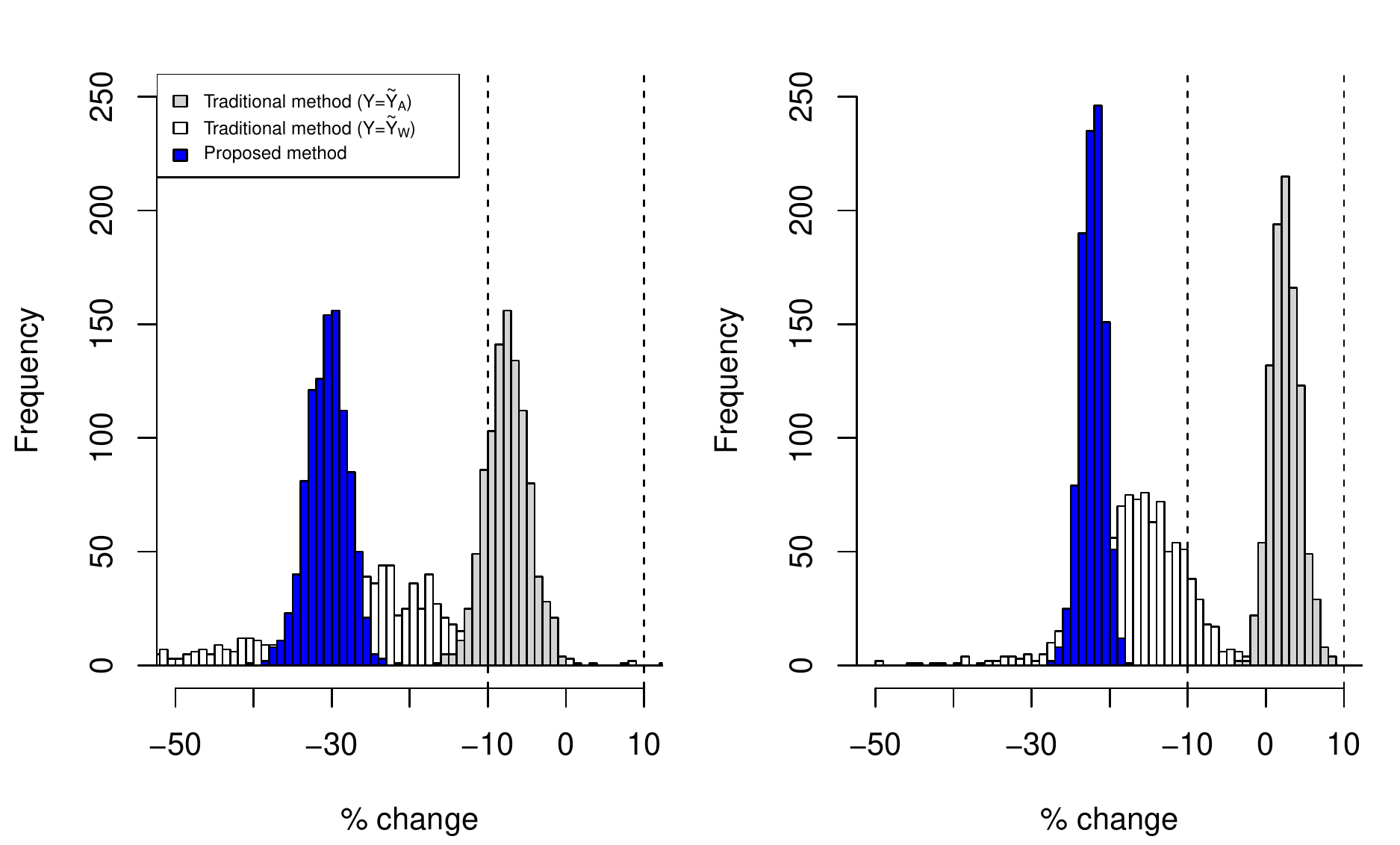}
	\caption{Percentage change of the association after adjusting for risk factors in the analysis of 1000 synthetic data sets. Considered are two settings: (a) mimicking the real {\it well-being} data structure; and (b) doubling the coefficients $\hat{\beta}_A$  the fitted model $Y_W \sim Y_A+\bm X$. The vertical dashed lines indicate the thresholds of a clinically significant moderation effect. } \label{fig:simu1}
\end{figure}

For each synthetic well-being data set, we calculate the percentage change of the association between $\tilde{Y}_{W}$ and $\tilde{Y}_{A}$ after adjusting for $\tilde{\bm X}$, namely, 
\[\frac{\mathcal{T}(\tilde{Y}_{W}, \tilde{Y}_{A} : \tilde{\bm X}) - \mathcal{T}(\tilde{Y}_{W}, \tilde{Y}_{A})}{\mathcal{T}(\tilde{Y}_{W}, \tilde{Y}_{A})} \times 100\%.
\]
Similar to the real data analysis in Table \ref{tab:ex1}, we compare in Figure~\ref{fig:simu1} the results from our proposed method (blue) and the traditional regression method with two choices $\tilde{Y}_{W} \sim \tilde{Y}_{A}+ \tilde{\bm X}$ (white) and $\tilde{Y}_{A} \sim \tilde{Y}_{W}+ \tilde{\bm X}$ (grey). The three histograms in Figure \ref{fig:simu1}(a) depict the distributions of the percentage changes in association, which visually describe the impact of the moderation effect of the risk factors $\tilde{\bm X}$. When $\tilde{Y}_{W}$ is treated as the response and $\tilde{Y}_{A}$ a predictor, the white histogram in Figure \ref{fig:simu1}(a) has a mean value of $-30.2\%$ and it spreads out wildly (SD=$59.9\%$). When the roles of $\tilde{Y}_{W}$ and $\tilde{Y}_{A}$ are switched in the regression, the grey histogram shifts much closer to 0 with the mean value being $-7.5\%$ and a much narrower range (SD=$2.7\%$). The substantial difference between the two histograms confirms that the disparity observed in real data analysis ($-34.5\%$ versus $-9.3\%$) is not an exception but the norm for the traditional regression coefficient method. The choice of using which as the response or predictor may result in very different sizes of the moderation effect. In contrast, our method yields the blue histogram with a mean of $-30.4\%$ and SD=$2.6\%$, which is consistent with the percentage change of $-34.0\%$ in real data analysis.

We modify the simulation setting by doubling $\hat{\beta}_A$ in the fitted model $Y_W \sim Y_A+\bm X$ (i.e., $2\times\{-2.180, -7.341, -15.526, -23.446\})$. As a result, the strength of the underlying partial association is increased and the size of moderation effect is decreased. In this situation, Figure \ref{fig:simu1}(b) shows that the regression  coefficient method generates even more confusing result. The white histogram ($\tilde{Y}_{W} \sim \tilde{Y}_{A}+ \tilde{\bm X}$) contains mostly negative values with a mean of $-20.5\%$, whereas the grey histogram ($\tilde{Y}_{A} \sim \tilde{Y}_{W}+ \tilde{\bm X}$) sits on the positive side with a mean of $2.5\%$. Considering that $10\%$ is often used as a threshold to determine if the moderation effect is clinically significant, the regression coefficient method leads to completely different (and contradictory) conclusions --- the moderation effect of the risk factors is (or {\it is not}) clinically significant if the regression $\tilde{Y}_{W} \sim \tilde{Y}_{A}+ \tilde{\bm X}$ (or $\tilde{Y}_{A} \sim \tilde{Y}_{W}+ \tilde{\bm X}$) is used to draw inference. Our simulation results demonstrate that the choice of the response or predictor may not only change the effect size as seen in Figure \ref{fig:simu1}(a), but also alter the sign or direction of the effect as seen in Figure \ref{fig:simu1}(b). In contrast, our proposed method yields a (blue) histogram that has a mean of $-22.2\%$ (SD=$1.5\%$), which indicates a moderation effect that is statistically and clinically significant. The effect size $22.2\%$ in (b) is smaller than $30.4\%$ in (a), which is consistent with the way the simulation setting is modified with a lessened moderation effect.

To summarize, our simulation results, coupled with the real data results in the introduction, provide numerical evidence that the traditional regression coefficient  method may deliver inconsistent results and lead to contradictory conclusions. We recommend using a correlation-like measure (such as our $\mathcal{T}$-measure), in place of regression coefficients, to size the moderation effect, particularly when the two outcomes have a bi-directional cause/effect relationship.

\subsection{Visualizing partial association}
\begin{figure}
	\centering 		
	\includegraphics[scale=0.50]{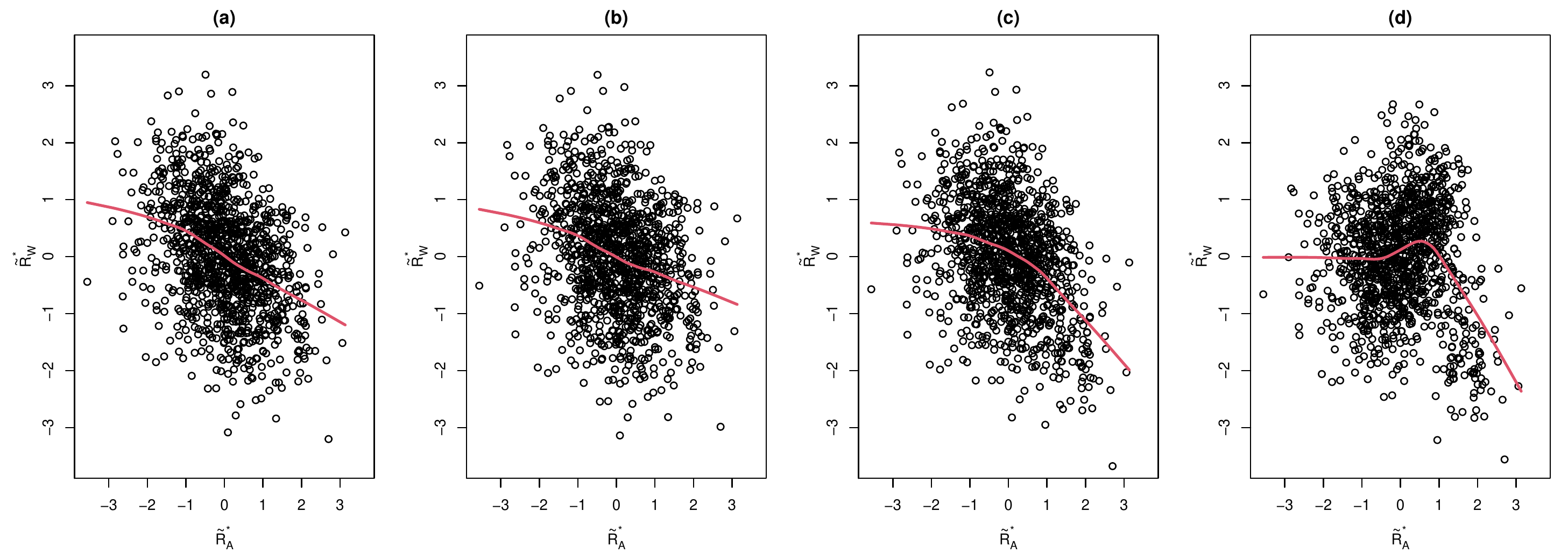}
	\caption{Partial regression plots of the synthetic outcome variables {\it well-being} and {\it anxiety} under the four settings described in Section 3.2. Red-colored curve is the fitted smooth curve using the LOWESS method.} \label{fig:simuvisual}
\end{figure}

Another advantage of our method is that it allows us to visualize partial association and potentially capture non-linear association structures. To demonstrate it, we consider four simulation settings by varying the coefficient $\hat{\beta}_A$ for $\tilde{Y}_{A}$ in generating $\tilde{Y}_{W} \sim \tilde{Y}_{A}+ \tilde{\bm X}$. They are (a) using the estimate $\hat{\beta}_{A}=(-2.2, -7.3, -15.5, -23.4)$ from real data analysis; (b)  setting a linear partial relationship by using $\hat{\beta}_{A}=(-2, -7, -12, -17)$; (c) setting a non-linear but monotonic relationship by using $\hat{\beta}_{A}=(-2, -3, -10, -30)$; and (d) setting a non-monotonic relationship by using $\hat{\beta}_{A}=(-2, -3, 10, -30)$. We visualize the partial association between $\tilde{Y}_{W}$ and $\tilde{Y}_{A}$ using partial regression plots as illustrated in Figure \ref{fig:simuvisual}.  

The trend captured by the LOWESS curve in each of the plots in Figure~\ref{fig:simuvisual} is consistent with the underlying partial relationship set in the simulation of the synthetic data. For example, the LOWESS curves in Figure \ref{fig:simuvisual}(a) and (b) are similar, and both partial relationships seem to be  approximately linear. A nonlinear but monotone partial relationship is revealed by Figure \ref{fig:simuvisual}(c), which is consistent with the nonlinear reduction in the sequence of the four coefficients in $\hat{\beta}_{A}=(-2, -3, -10, -30)$. Figure \ref{fig:simuvisual}(d) shows that our partial regression plot can even capture a non-monotonic relationship. The LOWESS curve is almost flat on the left but sharply declines on the right, and in between there is a bump. This pattern correctly reflects the way we set up $\hat{\beta}_{A}=(-2, -3, 10, -30)$, which contains a relatively ``flat'' component $\{-2, -3\}$, a sharp decline $\{-30\}$, and a ``bump'' $\{10\}$ in between. In addition to our numerical measures $\{-0.26, -0.20, -0.24, 0.007\}$ for (a)-(d), our partial regression plots allow us to carry out further inspection by depicting association structures. Our visualization tool therefore provides an important complement to the quantitative assessment. 

%

\subsection{Testing partial dependence}

Our last investigation is on statistical power for detecting partial dependence. We benchmark our method against two existing methods: the gold standard likelihood ratio test (LRT) and the nonparametric covariate-adjusted association test (NCAT) by \citet{zhu2012nonparametric}. We consider the following data generating procedure. Let $Y_1$ be a five-level ordinal variable that follows the adjacent category logit model $\log \frac{P(Y_1=j)}{P(Y_1=j+1)}=\alpha_j+\bm \beta_1^\prime \bm X$ for $j=1,...,4$, where $\alpha_j=(-3, -2,  0, 2)$ and $\bm \beta_1=(-0.5, 1.5)^\prime$. The covariate vector $\bm X=(X_1, X_2)^\prime$ with $X_1 \sim N(0, 2^2)$ and $X_2 \sim U(0, 1)$. The continuous outcome $Y_2$ is generated from the linear model $Y_2=\eta_1 \mathbb{I}(Y_1=1)+\eta_2 \mathbb{I}(Y_1=2)+ \dotso+ \eta_5 \mathbb{I}(Y_1=5)+ \bm \beta_2^T \bm X + \epsilon$, where $\bm \beta_2=(1, 1.5)^\prime$. The coefficient vector $\bm \eta=(\eta_1,...,\eta_5)$ controls the partial association between $Y_1$ and $Y_2$. We consider (a) linear $\bm \eta=\lambda\times (0, 1, 2, 3, 4)$, (b) quadratic $\bm \eta=[\lambda\times (0, 1, 2, 3, 4)]^2$, and (c) exponential $\bm \eta=\exp[\lambda\times (0, 1, 2, 3, 4)]$ partial associations. We vary the value of $\lambda$ to represent different levels of  association strength. When $\lambda=0$, $Y_1$ and $Y_2$ are partially independent. For each setting, we generate 1000 independent samples with a sample size of 200.

\begin{figure}
	\centering 		
	\includegraphics[scale=0.46]{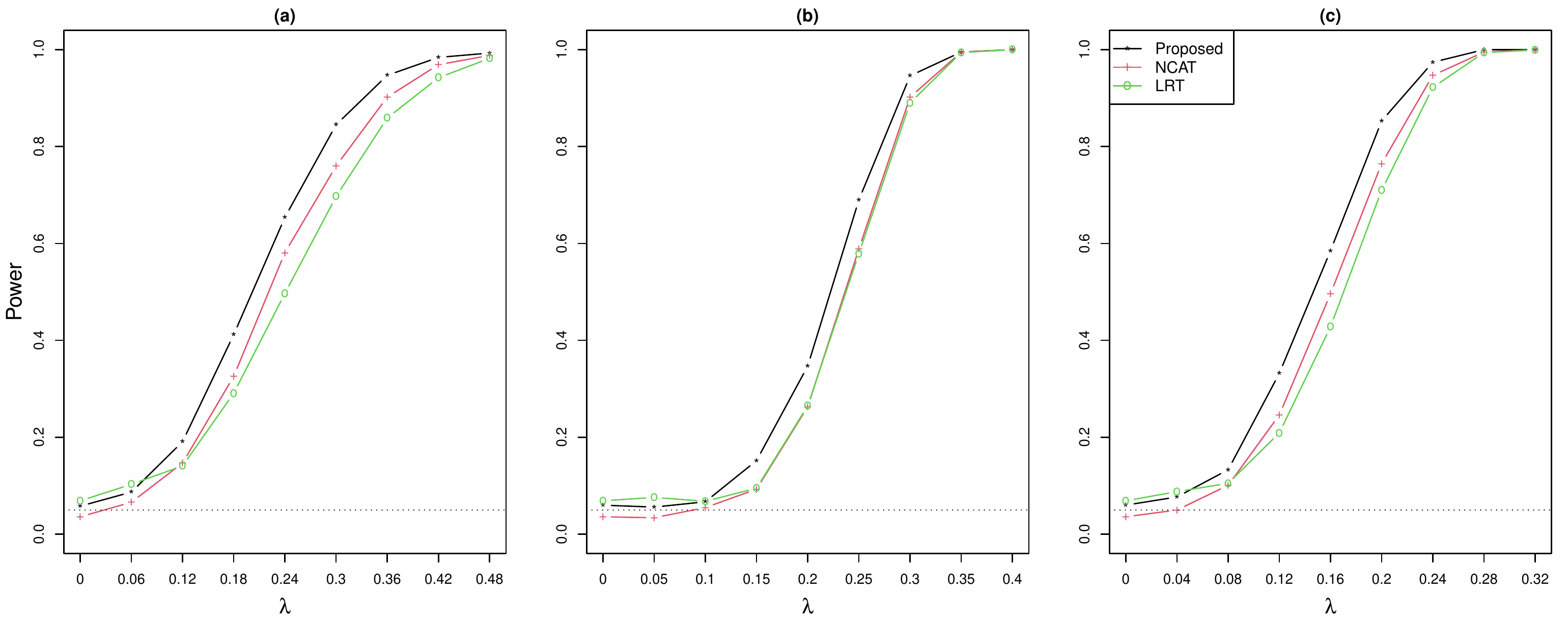}
	\caption{Statistical power of the proposed, NCAT, and LRT methods for testing partial independence between a continuous outcome and an ordinal outcome. Considered are (a) linear, (b) quadratic, and (c) exponential partial associations. The value of $\lambda$ represents the association strength, and it varies across the horizontal axis.} \label{fig:power}
\end{figure}

We test the null hypothesis of $Y_1$ and $Y_2$ being partially independent while adjusting for $\bm X$. The type I error and testing power of the three methods are compared in Figure \ref{fig:power}. The value of $\lambda$ varies across the horizontal axis. When $\lambda=0$, all the three methods yield type I errors that are close to 0.05 across the settings (a)-(c). When $\lambda$ increases away from zero, our method appears to have the highest power to detect partial dependency. Note that the LRT method tests $\eta_1=\eta_2=\cdots=\eta_5=0$ for five parameters simultaneously. It thus has the multiple testing problem, which may have undermined its power. Our method consolidates partial association information into a single parameter and tests $\mathcal{T}=0$. This difference may explain why our method outperforms the LRT which is often deemed as the gold standard.

\section{Discussion}\label{sec:summary}

Motivated by the analysis of the COVID-19 impact on college students, our paper has solved a statistical challenge in dealing with mixed data in the YOU survey. We map the residual randomness (after adjusting for covariates) to the same continuous scale, regardless of the outcomes being binary, ordinal, or continuous. It is achieved by defining a unified residual that broadly applies to mixed data and a wide range of regression models. As a result, the method allows us to adjust for covariates for multivariate outcomes by using a possibly mixture of linear or non-linear, cumulative or non-cumulative link models. It substantially broadens the scope and applicability of the surrogate method used in \citep{liu2018residuals, cheng2021surrogate, liu2021assessing, liu2022new}, which nevertheless focused on discrete data with a latent variable structure. Our development in this paper shows that the surrogate method can be used without relying on a latent variable structure. It therefore has advanced the theory of the surrogate method. 

For the study of partial association, our framework provides a unifying analysis in the following three aspects: (i) it unifies the analysis of different types of outcome variables, be it continuous, binary, or ordinal; (ii) it unifies the analysis using different regression models (e.g., cumulative or non-cumulative link models) for covariate adjustments; and (iii) it unifies a numerical measure, a testing procedure for simple and composite hypotheses, and a visualization tool in the same framework. As a result of (i) and (ii), analysis results can be compared to each if our framework is applied to different types of data or models. As a result of (iii), our $\mathcal{T}$ measure, $p$-value, and partial regression plot can be used to strengthen and complement each other. The analysis using our method has revealed the COVID-19 impact in a more complex three-way relationship. The risk factors collected in the surveys have been found to have a significant moderation effect in reducing the correlation between students' well-being and other mental health measures. In particular, the risk factors {\it physical healthiness}, {\it loneliness}, and {\it accommodation} played a much stronger role in the 2020 student cohort in terms of moderating the association between {\it well-being} and {\it anxiety}. This finding has important implications for our post-COVID life when remote learning will likely become common in various forms. University administrators should take actions to develop policies and programs to help college students improve their physical health, social connections, and learning and living environments in the reshaped college life.

The challenge posed by mixed data is not unique to our analysis of the YOU survey. It is common in surveys that address subjects' well-being or mental health issues. The metrics used often yield quantitative (e.g., numerical scores) as well as qualitative (e.g., yes/no or 5-category ratings) outcomes. To study their multivariate relationships, this paper provides a numeric measure and a visualization tool in addition to merely a $p$-value for testing the simple hypothesis. The combination of these analysis tools can be used in routine exploratory/descriptive analysis of survey data in general. The results in various forms (e.g., our $\mathcal{T}$-measure, partial regression plot, and $p$-value) may likely provide useful insights before carrying out more sophisticated tasks such as modeling/clustering multivariate mixed data, which is of further research interest.


\bibliographystyle{asa}

\end{document}